\documentclass[preprint,prb,aps,epsf,eqsecnum,tightenlines]{revtex4}
\begin{document}

\title{New Developments in the Eight Vertex Model II. Chains of odd length}

\author{Klaus Fabricius
\footnote{e-mail Fabricius@theorie.physik.uni-wuppertal.de}}
\affiliation{ Physics Department, University of Wuppertal, 
42097 Wuppertal, Germany}
\author{Barry~M.~McCoy
\footnote{e-mail mccoy@insti.physics.sunysb.edu}}               
\affiliation{ Institute for Theoretical Physics, State University of New York,
 Stony Brook,  NY 11794-3840}
\date{\today}
\preprint{YITPSB-04-49}


\begin{abstract}
We study the transfer matrix of the 8 vertex model with an odd
number of lattice sites $N.$
For systems at the root of unity points $\eta=mK/L$ with $m$  odd
the transfer matrix is known to satisfy the famous ``$TQ$'' equation 
where ${\bf Q}(v)$ is a specifically known matrix. 
We demonstrate that the location of the zeroes of this ${\bf Q}(v)$ matrix
is qualitatively different from the case of even $N$ and in particular
they satisfy a previously unknown equation which is more general than
what is often called ``Bethe's equation.'' For the case of even $m$
where no ${\bf Q}(v)$ matrix is known we demonstrate that there are
many states which are not obtained from the formalism of the SOS model
but which do satisfy the $TQ$ equation.
The ground state for the particular case of $\eta=2K/3$ and
$N$ odd is investigated in detail.
\end{abstract}

\maketitle

\section{Introduction}
\label{intro}

The eigenvalues of the transfer matrix  ${\bf T}(v)$ of the eight vertex model
with periodic boundary conditions were computed long ago
by Baxter \cite{bax71},\cite{bax72}.
This transfer matrix, as given in \cite{bax72} by
\begin{equation}
{\bf T}(u)|_{\mu,\nu}={\rm Tr}W(\mu_1,\nu_1)W(\mu_2,\nu_2)\cdots W(\mu_N,\nu_N)
\end{equation}
where $\mu_j,\nu_j=\pm1$ and $W(\mu,\nu)$ is a $2\times 2$ matrix whose
nonvanishing elements are given as
\begin{eqnarray}
W(1,1)|_{1,1}&=W(-1,-1)|_{-1,-1}=\rho\Theta(2\eta)\Theta(v-\eta)H(v+\eta)=a(v)
\nonumber\\
W(-1,-1)|_{1,1}&=W(1,1)|_{-1,-1}=\rho\Theta(2\eta)H(v-\eta)\Theta(v+\eta)=b(v)
\nonumber\\
W(-1,1)|_{1,-1}&=W(1,-1)|_{-1,1}=\rho H(2\eta)\Theta(v-\eta)\Theta(v+\eta)=c(v)
\nonumber\\
W(1,-1)|_{1,-1}&=W(-1,1)|_{-1,1}=\rho H(2\eta)H(v-\eta)H(v+\eta)=d(v),
\label{bw}
\end{eqnarray}
has the important property that is satisfies the
commutation relation
\begin{equation}
[{\bf T}(v),{\bf T}(v')]=0.
\end{equation}
The definition and some useful properties of $H(v)$ and $\Theta(v)$
are summarized in  appendix A.

The solution the 1972 paper \cite{bax72} is restricted to the root of
unity condition
\begin{equation}
\eta=mK/L
\label{root72}
\end{equation}
and a key property of the computation of the eigenvalues of ${\bf T}(v)$ is the
definition of an
auxiliary matrix ${\bf Q}(v)$ with
the commutation properties that
\begin{equation}
[{\bf T}(v),{\bf Q}(v')]=0
\label{tqcom}
\end{equation}
\begin{equation}
[{\bf Q}(v),{\bf Q}(v')]=0\label{qqcom}
\end{equation}
and which satisfies the ``TQ'' equation
\begin{equation}
{\bf T}(v){\bf Q}(v)=[\rho \Theta(0)h(v-\eta)]^N{\bf Q}(v+2\eta)
+[\rho \Theta(0)h(v+\eta)]^N{\bf Q}(v-2\eta)
\label{tq}
\end{equation}
where
\begin{equation}
h(v)=\Theta(v)H(v).
\end{equation}

The matrix ${\bf Q}_{72}(v)$ defined in appendix C of \cite{bax72} 
was found in \cite{fm} (which will be referred to as paper I)  
to have the
quasi-periodicity properties
\begin{eqnarray}
{\bf Q}_{72}(v+2K)&={\bf SQ}_{72}(v)\label{qp721}\\
{\bf Q}_{72}(v+2iK')&=q^{-N}e^{-iN\pi v/K}{\bf Q}_{72}(v)
\label{qp722}
\end{eqnarray}
where
\begin{equation}
{\bf S}=\prod_{j=1}^N\sigma^z.
\end{equation}
The operator $S$ commutes with ${\bf T}(v)$ and ${\bf Q}_{72}(v)$
\begin{equation}
[{\bf T}(v),{\bf S}]=[{\bf Q}(v),{\bf S}]=0
\end{equation}
and has eigenvalues $\pm 1$. Therefore we may diagonalize ${\bf Q}_{72}(v)$
in the
space in which ${\bf S}$ is diagonal and in this subspace we see from
(\ref{qp721}) and (\ref{qp722}) that the eigenvalues
$Q_{72}(v)$ of ${\bf Q}_{72}(v)$ satisfy
\begin{eqnarray}
Q_{72}(v+2K)&=(-1)^{\nu'}Q_{72}(v)\label{qp1}\\
Q_{72}(v+2iK')&=q^{-N}e^{-iN\pi v/K}Q_{72}(v).\label{qp2}
\end{eqnarray}

We note that under spin inversion $\sigma^z_j\rightarrow -\sigma^z_j$
the eigenvalues of the transfer matrix are  invariant but that
${\bf S}\rightarrow (-)^N{\bf S}$. When $N$ is odd each eigenvalue of
${\bf T}(v)$ is therefore doubly degenerate; 
one with $\nu'=0$ and one with $\nu'=1$. This phenomenon does not happen
for $N$ even.

We further found in paper I \cite{fm} that for even  
$N$ if $m$ is even and $N\geq L-1$ then
${\bf Q}_{72}(v)$
does not exist and that for this case the eigenvalues of ${\bf T}(v)$
could only be computed from (\ref{tq}) by use of the symmetry property 
of the eigenvalues of the transfer matrix $T(v;\eta)$ 
valid for $N$ even
\begin{equation}
T(v+K;K-\eta)=(-1)^{\nu'}T(v;\eta)
\end{equation}
Thus for even $N$ when the root of unity condition (\ref{root72})
holds all eigenvalues of ${\bf T}(v)$ may be studied by
means of the matrix ${\bf Q}_{72}(v)$ alone.

An alternative method of computing the eigenvalues of ${\bf T}(v)$
which gives in addition the
eigenvectors of the transfer matrix was presented by Baxter in
1973 \cite{bax731}-\cite{bax733}. In the course of this computation in sec.
 6 of \cite{bax731} a new matrix  ${\bf Q}(v)$ is defined for even $N$
only (for all $\eta$ not just those satisfying (\ref{root72}))
which also has the properties
(\ref{tqcom})-(\ref{tq}) found in \cite{bax72}. 
However, unlike the matrix of the 1972 construction, this new matrix
commutes with the spin reflection operator ${\bf R}$ which sends
the indices $\mu_j$ and $\nu_j$ into their negatives and instead of
the quasiperiodicity conditions (\ref{qp721}) and (\ref{qp722}) satisfies
the quasiperiodicity conditions (10.5.36a) and (10.5.43.a) of \cite{baxb}
\begin{eqnarray}
{\bf Q}_{73}(v+2K)&={\bf SQ}_{73}(v)\label{qp731}\\
{\bf Q}_{73}(v+iK')&={\bf RS}q^{-N/4}e^{-iN\pi v/2K}{\bf Q}_{73}(v).
\label{qp732}
\end{eqnarray}

In this paper we extend the considerations of paper I to odd $N$
and contrast the construction of the matrix ${\bf Q}_{72}(v)$ of 
\cite{bax72} with the construction of eigenvalues and eigenvectors of
the transfer matrix ${\bf T}(v)$ of \cite{bax731}-\cite{bax733}.

For the case of $m$ odd in \ref{root72} the matrix ${\bf Q}_{72}(v)$ defined in \cite{bax72}
exists. In sec.2 where we numerically compute the 
zeroes of the eigenvalues ${\bf Q}_{72}(v)$ for odd $N$ and
find that they are qualitatively different
from the case of $N$ even.
We also consider the construction of eigenvectors of
\cite{bax731}-\cite{bax733}.
However this method of computation of eigenvectors
requires
((4.2) of \cite{bax732} and (1.11) of \cite{bax733}) that
in order to satisfy the periodic boundary conditions
the number of sites $N$ must be of the form
\begin{equation}
N=2n_B+L_{73}n_L
\label{cond73}
\end{equation}
where $L_{73}$ is defined from
\begin{equation}
\eta=2m_{73}K/L_{73}.
\label{root73}
\end{equation}
When $m$ in the root of unity condition (\ref{root72})
is odd we see that (\ref{cond73}) becomes
\begin{equation}
N=2n_B+2Ln_L
\label{cond732}
\end{equation}
which can obviously not hold for odd $N$ and thus the methods of
\cite{bax731}-\cite{bax733} are not applicable.

For odd $N$ when $m$ is even or when $\eta$ is not of the form (\ref{root72})
there is no analytic proof of the existence of a matrix ${\bf Q}(v)$
with the properties (\ref{tqcom})-(\ref{tq}). Nevertheless since every
irrational number can be well approximated by a rational number of the
form (\ref{root72}) it is of interest to
consider (\ref{tq}) as an equation for eigenvalues and numerically
study the existence of solutions for the eigenvalues $Q(v)$ and to
compare these zeroes with the behavior of the zeros computed using the
explicit form of ${\bf Q}_{72}(v).$  
We do this
in sec. 3.

In sec. 4 we study the case where the root of unity condition
(\ref{root72}) holds with $m$ even. This case is of particular
interest because here some of the eigenvalues of the transfer matrix 
develop extra degeneracies over and beyond the double degeneracy of
all eigenvalues which exists for odd $N$ for all values of $\eta$
because of spin reflection invariance. 
Furthermore in this case,  even though no matrix ${\bf Q}(v)$ is known, 
the methods of \cite{bax731}
-\cite{bax733} allow the computation of (at least some )
eigenvectors and eigenvalues of ${\bf T}(v)$
provided that (\ref{cond73}) is satisfied
which for even $m$ requires that
\begin{equation}
N=2n_B+Ln_L
\label{res73}
\end{equation}
with $n_B$ and $n_L$ integers.
We will see that $n_B$ is the number of pairs of roots
\begin{equation}
v^B_k {\rm and}~ v^B_k+iK'
\label{bethe}
\end{equation}
 of the $Q(v)$ which solves the $TQ$ equation
with $T(v)$ and $Q(v)$ considered as scalars and that $n_L$ is the
number of $L$-strings of the form
\begin{equation}
v^L_k+2jK/L,~~~{\rm with}~j=0,\cdots, L-1.
\label{strings}
\end{equation}
We study these scalar solutions
numerically  and find 
that while there are solutions for $Q(v)$
which do
have paired roots and L-strings there are also solutions with $N$ roots where
there are neither paired roots nor $L$-strings.
The solutions of $Q(v)$  with neither paired roots nor $L$-strings
are not consistent with the restriction (\ref{res73}).
We therefore conclude that for odd $N$ there are eigenvectors of the
transfer matrix ${\bf T}(v)$ which cannot be obtained by the methods of
\cite{bax731}-\cite{bax733}.
These eigenvectors are significant because they include
the ground state of the XYZ spin chain.
We study this case analytically for $\eta=2K/3$ in sec. 5.

In sec. 6 we study the 6 vertex limit of the 8 vertex model where
the associated spin chain becomes
\begin{equation}
H_{XXZ}=-{1\over 2}\sum_{j=1}^N\{\sigma_j^x\sigma_{j+1}^x+
\sigma_{j}^y\sigma_{j+1}^y+\Delta\sigma_j^z\sigma_{j+1}^z\}
\label{hxxz}
\end{equation}
with $\Delta=\cos \pi m/L,$
$\sigma^k_j$ are Pauli spin matrices at
site $j$ and periodic boundary conditions are imposed.
When $m$ is odd we  find that for every $Q(v)$ with $n<N/2$
zeros which satisfies
the $TQ$ equation (\ref{tq}) there is a second solution with $N-n$ zeros
which also satisfies (\ref{tq}) with the same eigenvalue of ${\bf T}(v)$.
This is the extension to rational values of
$\gamma$ of the phenomenon
discovered by Pronko and Stroganov \cite{ps} for irrational $\gamma$
 and by Bazhanov, Lukyanov and Zamolodchikov \cite{blz1}, \cite{blz2}
in the context of conformal field theory.
We contrast our results with the case which has been previously studied
\cite{bax2002}, \cite{dfm}-\cite{fm4} of rational $\gamma$ with $N$ even
and also make contact with the special properties  which  the value
$\Delta=-1/2$ has for odd $N$ \cite{strog},\cite{rs},\cite{dbnm}.
We conclude with a discussion of the significance of our results in sec. 7.

\section{Odd N for $\eta=mK/L$ with m odd and L even or odd}

When $N$ is odd with $m$  odd and $L$  even or odd then
the restriction (\ref{cond732}) cannot be
satisfied but the matrix
${\bf Q}_{72}(v)$ exists. Thus in this case we may proceed
as we did in the case of even $N$ in paper I \cite{fm} and 
numerically determine the zeroes of the eigenvalues of 
${\bf Q}_{72}(v)$ directly from the definition of \cite{bax72}.

The most general function which satisfies the
quasi-periodicity properties (\ref{qp1}) and (\ref{qp2})
can be written in the factorized form
\begin{equation}
Q_{72}(v)={\cal K}(q,v_k){\rm exp}(-i \nu \pi v/2K)\prod_{j=1}^{N}
H(v-v_j)
\label{fac3}
\end{equation}
where the $v_j$ are unique once we adopt a convention for
the location of the fundamental region of the quasiperiodic function $H(v).$
Substituting (\ref{fac3}) in (\ref{qp1}) and using (\ref{p1}) we find
\begin{equation}
e^{i\pi (\nu'+\nu+N)}=1~~{\rm so}~~\nu'+\nu+N={\rm even~integer}
\label{qp1a}
\end{equation}
and substituting (\ref{fac3}) in (\ref{qp2}) and using (\ref{p2}) we find
\begin{equation}
e^{\pi i(-i\nu K'/K+N+\sum_{j=1}^N v_j /K)}=1
~~{\rm so}~~N+(-\nu iK'+\sum_{j=1}^Nv_j)/K={\rm even~integer}
\label{qp2a}.
\end{equation}
Taking real and imaginary parts of (\ref{qp2a}) we obtain the sum rule
\begin{equation}
N+\sum_{j=1}^N {\rm Re} (v_j)/K= {\rm even~ integer}
\label{sumrule}
\end{equation}
and  find that $\nu$ satisfies
\begin{equation}
\nu=\sum_{j=1}^N{\rm Im} (v_j)/K'={\rm even~ integer} -\nu'-N.
\label{nuform}
\end{equation}
where the value of the even integer and thus the numerical value of $\nu$
depends on the conventions used to specify the fundamental region for $v_j.$

We have numerically studied the zeros of all the eigenvalues
$Q_{72}(v)$. For the case of even
$N$ we found in paper I \cite{fm} that the zeros either occur in pairs
(\ref{bethe}) which we call Bethe roots or $L$-strings (\ref{strings}).
Therefore for even $N$ all the eigenvalues  $Q_{72}(v)$ may be written
as
\begin{eqnarray}
&Q_{72}(v)={\cal K}(q,v_k){\rm exp}(-i \nu \pi v/2K)
\prod_{j=1}^{n_B}
H(v-v_j^B)H(v-v_j^B-iK')\nonumber\\
&\times \prod_{j=1}^{n_L}H(v-v^L_j)
H(v-v^L_j-2K/L)\cdots H(v-v^L_j-2(L-1)K/L)
\label{fac1}
\end{eqnarray}
where
\begin{equation}
2n_B+Ln_L=N.
\end{equation}
Using (\ref{thtrans}) we may rewrite this as
\begin{eqnarray}
&Q_{72}(v)={\tilde {\cal K}}(q,v_k)
{\rm exp}(-i (\nu-n_B) \pi v/2K)(-i)^{n_B}q^{-n_B/4}\prod_{j=1}^{n_B}
h(v-v^B_j)\nonumber\\
&\times \prod_{j=1}^{n_L}H(v-v^L_j)H(v-v^L_j-2K/L)\cdots
H(v-v^L_j-2(L-1)K/L).
\label{fac2}
\end{eqnarray}
where
\begin{equation}
{\tilde {\cal K}}(q,v_k)={\cal K}(q,v_k)e^{-\pi i \sum_{j=1}^{n_B}v_j^B/(2K)}
\end{equation}

The roots $v^B_j$ are called  Bethe roots and by substitution of
(\ref{fac2}) into the $TQ$ equation (\ref{tq}) with $v$ set equal
to $v^B_j$ we see that the $v^B_j$ satisfy the ``Bethe's''
equation
\begin{equation}
\left({h(v^B_l-\eta)\over h(v_l^B+\eta)}\right)^N
=e^{2\pi i (\nu-n_B) m/L}\prod^{n_B}_{{j=1\atop j\neq l}}
{h(v_l^B-v_j^B-2\eta)\over h(v_l^B-v_j^B+2\eta)}
\label{bethe2}
\end{equation}

The roots $v_j^L$ are determined by the new
functional equation conjectured in I
\cite{fm}
\begin{equation}
{\bf A}'e^{-N\pi i v/2K}{\bf Q}_{72}(v-iK')
=\sum_{l=0}^{L-1}{h^N(v-(2l+1)\eta){\bf Q}_{72}(v)\over
{\bf Q}_{72}(v-2l\eta){\bf Q}_{72}(v-2(l+1)\eta)}
\label{conj}
\end{equation}
where ${\bf A}'$ is a matrix (called ${\bf A}^{-1}$ in ref. \cite{fm})
which commutes with ${\bf Q}_{72}(v),$ 
is independent of $v$ and depends on the
normalization in the construction of ${\bf Q}_{72}(v).$

The present case when $N$ is odd is quite different from the case $N$ even
previously studied. Now our numerical study finds that
there are neither the paired solutions (\ref{bethe})
nor the $L$ string solutions (\ref{strings}). Instead of the pairing condition
of roots (\ref{bethe}) we now find that there is a pairing of solutions
in the sense that for every set of roots $v_j$ there is a second set of
roots $v_j+iK'$ which also gives an eigenvalue of ${\bf Q}_{72}(v)$
and these two paired eigenvalues of ${\bf Q}_{72}(v)$ satisfy
the TQ equation with the same eigenvalue of ${\bf T}(v)$
Noting that
\begin{equation}
\sum_{j=1}^N {\bf Im}(v_j+iK')/K'=N+\sum_{j=1}^N Im(v_j)/K'
\end{equation} we conclude from (\ref{nuform}) that when $N$ is odd
the parity of $\nu'$ for the shifted solution is opposite to the
parity of $\nu'$ for the original solution. Thus since shifting
all roots by $iK'$ is equivalent to sending $v\rightarrow iK'$
we conclude that
\begin{equation}
Q_{72}(v+iK';\nu'=0)=Q_{72}(v;\nu'=1)
\label{qpair}
\end{equation}

We obtain an equation for the $N$ roots $v_j$ by using the
factorized form (\ref{fac3}) for $Q_{72}(v)$ and set $v=v_j$ in the TQ
equation (\ref{tq}) to obtain for odd $N$ with $m$ odd
\begin{equation}
\left({h(v_l-mK/L)\over h(v_l+mK/L)}\right)^N
=e^{2\pi i \nu m/L}\prod^{N}_{{j=1\atop j\neq l}}
{H(v_l-v_j-2mK/L)\over H(v_l-v_j+2mK/L)}.
\label{bethe3}
\end{equation}
which differs from the Bethe's equation  (\ref{bethe2}) for even $N$ in that
the function $H(v)$ appears on the right hand side instead of
$h(v)$ and the number of terms in the product is $N$ instead of $n_B.$
This equation seems to be new in the literature.

\section{Odd N and generic $\eta$}

For odd $N$ and $\eta$ not of the form (\ref{root72}) with $m$ odd
there is no analytic proof that there is a matrix 
${\bf Q}(v)$ with the properties (\ref{tqcom}) and (\ref{qqcom})
which satisfies the $TQ$ equation (\ref{tq}). 
However for finite $N$ if $L$ is sufficiently large it should be
impossible to tell the difference between values of $\eta$ which are
generic and those which satisfy (\ref{root72}) with $m$
odd. Consequently instead of computing the zeroes of $Q(v)$ using the
specific form of ${\bf Q}_{72}(v)$ given in ref. \cite{bax72}  
we may
consider (\ref{tq}) as an equation for eigenvalues, and see
numerically if for each eigenvalue $T(v)$ it is possible to find a
$Q(v)$ which satisfies (\ref{tq}). We have made such a study 
for $N=7$ and $N=9$ at $q=0.2$
and find
that for each eigenvalue of ${\bf T}(v)$ there do indeed exist two
distinct functions $Q(v)$ and $Q(v+iK')$ which satisfy the $TQ$
equation (\ref{tq}). These functions $Q(v)$ share with the eigenvalues
of ${\bf Q}_{72}(v)$ the property of having $N$ roots and these roots
satisfy the sum rules (\ref{sumrule}) and (\ref{nuform}) which follow
from the quasiperiodicity conditions 
\begin{eqnarray}
Q(v+2K)&=(-1)^{\nu'}Q(v)\label{nqp1}\\
Q(v+2iK')&=q^{-N}e^{-iN\pi v/K}Q(v).\label{nqp2}
\end{eqnarray}
which generalize the
quasiperiodicity conditions (\ref{qp1}) and (\ref{qp2}) of the
eigenvalues of ${\bf Q}_{72}(v).$ The $N$ roots $v_k$ satisfy the
generalization of (\ref{bethe3}) to generic values of eta  
\begin{equation}
\left({h(v_l-\eta)\over h(v_l+\eta)}\right)^N
=e^{2\pi i \nu \eta/K}\prod^{N}_{{j=1\atop j\neq l}}
{H(v_l-v_j-2\eta)\over H(v_l-v_j+2\eta)}.
\label{bethe4}
\end{equation}
We refer to (\ref{bethe4}) as the generic equation for roots.

We give in table I an example of the relation of
the zeroes of 2  eigenvalues of $Q(v)$ computed for $N=9$ at
$\eta=K/3$ from ${\bf Q}_{72}(v)$ and computed numerically at $\eta=.35K$ 
from the $TQ$ equation.

\newpage 

\begin{table}
\caption{A comparison of the roots of eigenvalues of ${\bf Q}_{72}(v)$
at $\eta=K/3$ with numerical solutions to the $TQ$ equation at
$\eta=.35K$ for the nome $q=0.20.$ The roots of state 2 and 4 are shifted
from those of state 1 and 3 by $iK' ({\rm mod}~ 2iK').$ 
States 3 and 4 are the two degenerate ground states.
The roots are
accurate to the number of significant figures given.}
\begin{tabular}{|l|l|l|}
State&$Q(v)~\eta=0.35K$&$Q_{72}(v)~\eta=K/3$ \\ \hline
1&$i0.06866174K'$&$i0.06598354K'$\\
&$i0.57652164K'$&$i0.57706877K'$\\
&$i0.83871494K'$&$i0.84250034K'$\\
&$i0.97126416K'$&$i0.97027666K'$\\
&$i1.08588486K'$&$i1.08005865K'$\\
&$i1.22542023K'+K$&$i1.21577690K'+K$\\
&$i1.48883198K'$&$i1.49156270K'$\\
&$i1.79747242K'$&$i1.80675767K'$\\
&$i1.94722804K'$&$i1.95008384K'$\\ \hline
2&$i1.06866174K'$&$i1.06598354K'$\\
&$i1.57652164K'$&$i1.57706877K'$\\
&$i1.83871494K'$&$i1.84250034K'$\\
&$i1.97126416K'$&$i1.97027666K'$\\
&$i0.08588486K'$&$i0.08005865K'$\\
&$i0.22542023K'+K$&$i0.21577690K'+K$\\
&$i0.48883198K'$&$i0.49156270K'$\\
&$i0.79747242K'$&$i0.80675767K'$\\
&$i0.94722804K'$&$i0.95008384K'$\\ \hline
3&$i0.09682452K'+K$&$i0.09803186K'+K$\\
&$i0.31149066K'+K$&$i0.31332654K'+K$\\
&$i0.56470429K'+K$&$i0.56389029K'+K$\\
&$i0.80075993K'+K$&$i0.79896112K'+K$\\
&$iK'+K$&$iK'+K$\\
&$i1.19924006K'+K$&$i1.20103887K'+K$\\
&$i1.43529570K'+K$&$i1.43610970K'+K$\\
&$i1.68850933K'+K$&$i1.68667345K'+K$\\
&$i1.90304175K'+K$&$i1.90196813K'+K$\\ \hline
4&$i1.09682452K'+K$&$i1.09803186K'+K$\\
&$i1.31149066K'+K$&$i1.31332654K'+K$\\
&$i1.56470429K'+K$&$i1.56389029K'+K$\\
&$i1.80075993K'+K$&$i1.79896112K'+K$\\
&$K$&$K$\\
&$i0.19924006K'+K$&$i0.20103887K'+K$\\
&$i0.43529570K'+K$&$i0.43610970K'+K$\\
&$i0.68850933K'+K$&$i0.68667345K'+K$\\
&$i0.90304175K'+K$&$i0.90196813K'+K$\\ \hline
\end{tabular}
\end{table}
\newpage

\section{Odd N for $\eta=m/L$ with m even and L odd}

When $\eta$ satisfies the root of unity condition (\ref{root72})
with $m$ even and when $N$ satisfies
$N=2n_B+Ln_L$
(some of) the eigenvectors and eigenvalues of the transfer matrix 
${\bf T}(v)$ may be computed by the methods of
\cite{bax731}-\cite{bax733}.
As with the case of generic $\eta$ there is no analytic proof 
in this case of the existence for odd $N$ of a matrix ${\bf Q}_o(v)$
which satisfies (\ref{tqcom})-(\ref{tq}) and to gain insight we
numerically solve the $TQ$ equation (\ref{tq}) for scalar functions
$Q(v)$ for (1) $\eta=2K/3$ and a nearby value 
$\eta=0.65K$ for $N=9$ and $q=0.2$ 
and for (2) $\eta=2K/5$ and a nearby value of $\eta=0.398$ for $N=7$
and $q=0.2$
For $\eta=0.65K$ and $0.398K$ the numerical solutions have 
the feature of generic
values of $\eta$ that they have $N$ roots which satisfy the sum rules
(\ref{sumrule}) and (\ref{nuform}). However, at precisely $\eta=2K/3$
and $\eta=2K/5$ 
there is qualitative change in the numerical solutions. 
For $\eta=2K/3$ there are only 2 solutions which have $N$ roots and
while for $\eta=2K/5$ for $N=7$ the number of solutions with $7$ roots
decreases from 128 to 58. All other solutions have less
than $N$ roots and they do not satisfy the sum rules (\ref{sumrule})
and (\ref{nuform}). We illustrate
this in table II and table IV for $\eta=2K/3$ and in tables V and VI
for $\eta=2K/5.$ 

What is clearly visible in the examples of tables II - V (and is seen in
all the rest of the  data) is that as $\eta \rightarrow 2mK/L$ two
phenomena occur:

1) There are $n_L L$ roots which move to form $n_L$ L strings where
   $L$ roots satisfy
\begin{equation}
\alpha_k,~\alpha_k+2\eta,\cdots, \alpha_k+2(L-1)\eta
\end{equation}
and make a contribution to $Q_o(v)$ of
\begin{equation}
\prod_{k=1}^{n_L}\prod_{j=0}^{L-1}H(v-\alpha_k-2j\eta)
\end{equation}
which cancels out of (\ref{bethe3}).

2) The remaining $n_B=(N-n_LL)/2$ roots arrange themselves in pairs $v_k$
   and $v_k+iK.'$

Taking into account both of these phenomena we see that the generic
equation for the roots (\ref{bethe3}) reduces to the ''Bethe's''
equation (\ref{bethe2}) derived in \cite{bax731}-\cite{bax733}.
However, the roots of those $Q_o(v)$ which do not have paired roots and
$L$ strings as $\eta\rightarrow 2mK/L$ satisfy the generic equation
(\ref{bethe3}) instead of the Bethe's equation (\ref{bethe2}).
We thus  conclude from these numerical studies that 
for odd $N$ with $m$ even there
exist eigenvalues of the transfer matrix which cannot be obtained by the
methods of \cite{bax731}-\cite{bax733}.
For these states the eigenvalues $Q_o(v)$ of the $TQ$  equation (\ref{tq}) 
have been found numerically to satisfy
\begin{equation}
\sum_{l=0}^{L-1}{h^N(v-(2l+1)\eta)\over
{Q}_o(v-2l\eta){Q}_o(v-2(l+1)\eta)}=0.
\label{conj2}
\end{equation}

If there is one $L$ string in the eigenvalue of ${\bf Q}_{o}(v)$ the
string center of the limiting values drops out of the $TQ$ equation but may
be computed from the $N-L$ roots which do satisfy the generic equation
(\ref{bethe4}) by means of the sum rule (\ref{nuform}).
When there are 3 or more $L$ strings equations for the string centers may be
obtained by again carefully taking the limit of the generic equation
(\ref{bethe4}) for the roots of $Q_o(v)$ by generalizing the
corresponding 6 vertex computation of ref. \cite{fm3} to find 
\begin{eqnarray}
&&\sum_{k=1}^{L}\frac{c_{0}^{-(k+1)}\Phi_{k+1}(\alpha_j)}{P_{k}(\alpha_j)
P_{k+1}(\alpha_j)}\left[
N\frac{h'(\alpha_{j}+(2k+1)\eta_{0})}{h(\alpha_{j}+(2k+1)\eta_{0})}\nonumber
-\sum_{i=1}^{n_{0}}\left(
\frac{H'(\alpha_{j}-v_{i}^{(0)}+(k+1)2\eta_{0})}
     {H(\alpha_{j}-v_{i}^{(0)}+(k+1)2\eta_{0})} +
\frac{H'(\alpha_{j}-v_{i}^{(0)}+k2\eta_{0})}
     {H(\alpha_{j}-v_{i}^{(0)}+k2\eta_{0})}\right)\right]\nonumber \\
&&-2K(\alpha_j)\sum_{l\neq j}\sum_{m=1}^{L}
\frac{H'(\alpha_{j}-\alpha_{l}+m 2\eta_{0})}
     {H(\alpha_{j}-\alpha_{l}+m 2\eta_{0})}  
+ K(\alpha_j)\frac{i\pi \nu}{K} = 0
\end{eqnarray}
where $v^{(0)}$ 
the ordinary Bethe roots which occur in pairs, $j=1,2,\cdots ,n_L,$
\begin{equation}
c_{0} = \exp(2\pi i\nu\eta/K)
\end{equation}
\begin{equation}
P_{k}(\alpha_j) = \prod_{i}^{n_{0}}H(\alpha_{j}-v_{i}^{(0)}+k2\eta_{0})
\end{equation}
\begin{equation}
\Phi_{k}(\alpha_j) = h^{N}(\alpha_{j}+(2k-1)\eta_{0})
\end{equation}
and
\begin{equation}
K(\alpha_j) = \sum_{k=1}^{L}\frac{c_{0}^{-(k+1)}\Phi_{k+1}(\alpha_j)}
{P_{k}(\alpha_j)P_{k+1}(\alpha_j)}
\end{equation}

\newpage

\begin{table}
\caption{A comparison of the roots of numerical solutions of the $TQ$
equation for $\eta=0.65K$ and $\eta=2K/3$ for $N=9$ and $q=0.2$. 
In states 1 and 2 a three string develops as $\eta \rightarrow 2K/3$
which drops out of the $TQ$ equation at $\eta=2K/3.$ At $\eta=2K/3$
the  six roots obtained from the $TQ$ equation occur in 
pairs $v_k$ and $v_k+iK'.$ States 3 and 4 are the two degenerate 
ground states which have no $L$ strings. 
The roots are
accurate to the number of significant figures given.}
\begin{tabular}{|l|l|l|}
State&$\eta=.65K$&$\eta=2K/3$ \\ \hline
1&$i0.1548264K'$&$i0.14488580K'$\\
&$i0.40419528K'$&$i0.39698346K'$\\
&$i0.92299060K'$&$i0.92807395K'$\\
&$i1.13102864K'$&$i1.14488580K'$\\
&$i1.37507482K'$&$i1.39698346K'$\\
&$i1.94364493K'$&$i1.92807395K'$\\
&$i1.68925695K'+0.27932034K$&\\
&$i1.68925695K'+1.72067964K$&\\
&$i1.68972529K'+K$&\\ \hline
2&$i1.1548264K'$&$i1.14488580K'$\\
&$i1.40419528K'$&$i1.39698346K'$\\
&$i1.92299060K'$&$i1.92807395K'$\\
&$i0.13102864K'$&$i0.14488580K'$\\
&$i0.37507482K'$&$i0.39698346K'$\\
&$i0.94364493K'$&$i0.92807395K'$\\
&$i0.68925695K'+0.27932034K$&\\
&$i0.68925695K'+1.72067964K$&\\
&$i0.68972529K'+K$&\\ \hline
3&$i0.06421637K'+K$&$i0.05992394K'+K$\\
&$i0.23183303K'+K$&$i0.21760180K'+K$\\
&$i0.61330441K'+K$&$i0.62586149K'+K$\\
&$i0.86288322K'+K$&$i0.89718639K'+K$\\
&$iK'+K$&$iK'+K$\\ 
&$i1.13711677K'+K$&$i1.12813602K'+K$\\
&$i1.38669558K'+K$&$i1.37413850K'+K$\\
&$i1.76816695K'+K$&$i1.78239820K'+K$\\ 
&$i1.93568362K'+K$&$i1.94007605K'+K$\\ \hline
4&$i1.06421637K'+K$&$i1.05992394K'+K$\\
&$i1.23183303K'+K$&$i1.21760180K'+K$\\
&$i1.61330441K'+K$&$i1.62586149K'+K$\\
&$i1.86288322K'+K$&$i1.89718639K'+K$\\
&$K$&$K$\\ 
&$i0.13711677K'+K$&$i0.12813602K'+K$\\
&$i0.38669558K'+K$&$i0.37413850K'+K$\\
&$i0.76816695K'+K$&$i0.78239820K'+K$\\ 
&$i0.93568362K'+K$&$i0.94007605K'+K$\\ \hline
\end{tabular}
\end{table}

\newpage

\begin{table}
\caption{Examples of the development of the octet of states with 3 $L$ strings
for $N=9$ as $\eta\rightarrow 2K/3$ with momentum  $P=-2\pi/3.$    
The numerical solutions of the
$TQ$ equation are given for $\eta=0.666K.$ 
The values of the corresponding $XYZ$ spin chain are shown 
to illustrate the approach to degeneracy.
Only 4 of the members of the octet are given. The remaining 4 are
obtained by sending $v_k\rightarrow v_k+iK'$ (mod $2iK'$).
The roots for corresponding state with $P=2\pi/3$ are the complex
conjugates (mod 2iK') of the roots for $P=-2K/3$
The roots are
accurate to the number of significant figures given.}
\begin{tabular}{|l|l|l|}
State&$\eta=.666K$&$H_{XYZ}$ \\ \hline 
1&$i0.10731398K'$&$-4.7526939$\\
 &$i0.10854909K'+.66600000K$&\\
 &$i0.10854909K'+1.3339999K$&\\
 &$i0.56699150K'+K$&\\
 &$i0.56696222K'+0.33190487K$&\\
 &$i0.56696222K'+1.66809512K$&\\
 &$i0.99234471K'$&\\
 &$i0.99120488K'+0.66600000K$&\\
 &$i0.99120488K'+1.33400000K$&\\ \hline
2&$i0.19439140K'+K$&$-4.7484930$\\
 &$i0.19440408K'+0.33199252K$&\\
 &$i0.19440408K'+1.66800747K$&\\
 &$i0.65737246K'$&\\
 &$i0.65549419K'+0.66600875K$&\\
 &$i0.65549419K'+1.33399124K$&\\
 &$i1.81540681K'$&\\
 &$i1.81651638K'+0.66600012K$&\\
 &$i1.81651638K'+1.33399987K$&\\ \hline
3&$i0.19872733K'+K$&$-4.7487136$\\
 &$i0.19874105K'+0.33199370K$&\\
 &$i0.19874105K'+1.66800629K$&\\
 &$i0.90240897K'$&\\
 &$i0.90240470K'+0.33199551K$&\\
 &$i0.90240470K'+1.66800448K$&\\
 &$i1.56554439K'+K$&\\
 &$i1.56551388K'+0.33190519K$&\\
 &$i1.56551388K'+1.66809480K$&\\ \hline
4&$i0.47511066K'$&$-4.7457518$\\
 &$i0.47701987K'+0.66597664K$&\\
 &$i0.47701987K'+1.33402335K$&\\
 &$i0.90529517K'+K$&\\
 &$i0.90291287K'+0.33199520K$&\\
 &$i0.90291287K'+1.66800480K$&\\
 &$i1.28586937K'$&\\
 &$i1.28455121K'+0.66599732K$&\\
 &$i1.28455121K'+1.33400267K$&
\end{tabular}
\end{table}

\newpage

\begin{table}
\caption{Examples of the development of the octet of states with 3 $L$ strings
for $N=9$ as $\eta\rightarrow 2K/3$ with momentum  $P=0.$    
The numerical solutions of the
$TQ$ equation are given for $\eta=0.666K.$ 
The values of the corresponding $XYZ$ spin chain are shown 
to illustrate the approach to degeneracy.
Only 4 of the members of the octet are given.. The remaining 4 are
obtained by sending $v_k\rightarrow v_k+iK'$ (mod $2iK'$).
The roots in states 1 and 2 are invariant under complex conjugations
(mod 2iK') and states 3 and 4 transform into each other under 
complex conjugation 
The roots are
accurate to the number of significant figures given.}
\begin{tabular}{|l|l|l|}
State&$\eta=.666K$&$H_{XYZ}$ \\ \hline 
1&$K$&$-4.7484601$\\
&$0.33199540K$&\\
 &$1.66800459K$&\\
 &$i0.69090441K'+K$&\\
 &$i0.69087039K'+0,33197194K$&\\
 &$i0.69087039K'+1.66802805K$&\\ 
&$i1.30955843K'+K$&\\
 &$i1.30912960K'+0.33197194K$&\\ 
 &$i1.30912960K'+1.66802805K$&\\ \hline
2&$K$&$-4.7440296$\\
 &$0.33199539K$&\\
 &$1.66800460K$&\\
 &$i0.39752759K'$&\\
 &$i0.39589406K'+0.66601414K$&\\
 &$i0.39589406K'+1.33398585K$&\\
 &$i1.60247240K'$&\\
 &$i1.60410593K'+0.66601414K$&\\
 &$i1.60410593K'+1.33398585K$&\\ \hline
3&$i0.30522748K'+K$&$-4.7511419$\\
 &$i0.30525934K'+0.33197051K$&\\
 &$i0.30525934K'+1.66802948K$&\\
 &$i0.78313587K'$&\\
 &$i0.78155612K'+0.66599948K$&\\
 &$i0.78155612K'+1.33400051K$&\\
 &$i1.91196073K'$&\\
 &$i1.91302248K'+0.66599999K$&\\
 &$i1.91302248K'+1.33400000K$&\\ \hline
4&$i1.69477251K'+K$&$-4.7511419$\\
 &$i1.69474065K'+0.33197051K$&\\
 &$i1.69474065K'+1.33802948K$&\\
 &$i1.21686441K'$&\\
 &$i1.21844389K'+0.66599948K$&\\
 &$i1.21844389K'+1.33400051K$&\\
 &$i0.08803926K'$&\\
 &$i0.086977515K'+0.66599999K$&\\
 &$i0.086977515K'+1.33400000K$&
\end{tabular}
\end{table}

\newpage

\begin{table}
\caption{A comparison of the roots of numerical solutions of the $TQ$
equation for $\eta=0.398K$ and $\eta=2K/5$ for $N=7$ and $q=0.2$
for states which do not develop $L$ strings or pairs $v_k$ and
$v_k+iK'.$ 
States 3 and 4 are the
two degenerate ground states.
 The roots are
accurate to the number of significant figures given.}

\begin{tabular}{|l|l|l|}
State&$\eta=.398K$&$\eta=2K/5$ \\ \hline
1&$i0.04397370K'$&$i0.04471961K'$\\
&$i0.20176882K'+K$&$i0.20257510K'+K$\\
&$i0.46160907K'$&$i0.46092396K'$\\
&$i0.81633633K'$&$i0.81538032K'$\\
&$i1.00694470K'$&$i1.00687865K'$\\
&$i1.59024152K'$&$i1.59035264K'$\\
&$i1.87912581K'$&$i1.87917007K'$\\ \hline
2&$i1.04397370K'$&$i1.04471961K'$\\
&$i1.20176882K'+K$&$i1.20257510K'+K$\\
&$i1.46160907K'$&$i1.46092396K'$\\
&$i1.81633633K'$&$i1.81538032K'$\\
&$i0.00694470K'$&$i0.00687865K'$\\
&$i0.59024152K'$&$i0.59035264K'$\\
&$i0.87912581K'$&$i0.87917007K'$\\ \hline
3&$i0.2553638K'+K$&$i0.25505369K'+K$\\
&$i0.58661122K'+K$&$i0.58678614K'+K$\\
&$i0.87889895K'+K$&$i0.87909858K'+K$\\
&$iK'+K$&$iK'+K$\\
&$i1.12110104K'+K$&$i1.12090141K'+K$\\
&$i1.41338877K'+K$&$i1.41321385K'+K$\\ 
&$i1.74463612K'+K$&$i1.74494630K'+K$\\ \hline
4&$i1.2553638K'+K$&$i1.25505369K'+K$\\
&$i1.58661122K'+K$&$i1.58678614K'+K$\\
&$i1.87889895K'+K$&$i1.87909858K'+K$\\
&$K$&$K$\\ 
&$i0.12110104K'+K$&$i0.12090141K'+K$\\
&$i0.41338877K'+K$&$i0.41321385K'+K$\\ 
&$i0.74463612K'+K$&$i0.74494630K'+K$\\ 
\end{tabular}
\end{table}

\newpage

\begin{table}
\caption{A comparison of the roots of numerical solutions of the $TQ$
equation for $\eta=0.398K$ and $\eta=2K/5$ for $N=7$ and $q=0.2$
for states which do develop 5 strings and paired roots.
The roots are
accurate to the number of significant figures given.}

\begin{tabular}{|l|l|l|}
State&$\eta=.398K$&$\eta=2K/5$\\ \hline
1&$i0.76120023K'$&$i0.76089181K'$\\
 &$i1.75808435K'$&$i1.76089181K'$\\
&$i1.48754767K'+K$&\\
&$i1.49676120K'+1.75698935K$&\\
&$i1.49676120K'+0.24301064K$&\\
&$i1.49982265K'+1.44611774K$&\\
&$i1.49982265K'+0.55388225K$&\\ \hline
2&$i1.76120023K'$&$i0.76089181K'$\\
 &$i0.75808435K'$&$i1.76089181K'$\\
&$i0.48754767K'+K$&\\
&$i0.49676120K'+1.75698935K$&\\
&$i0.49676120K'+0.24301064K$&\\
&$i0.49982265K'+1.44611774K$&\\
&$i0.49982265K'+0.55388225K$&\\ 
\end{tabular}
\end{table}

\section{The ground state for odd $N$ and $\eta=2K/3$}

To gain insight into these  solutions which have the pairing property
and are not obtained from \cite{bax731}-\cite{bax733} we consider
the simplest case
$\eta=2K/3$ and recall that in this case Baxter \cite{bax89} has,
by means of the
inversion relation found that the transfer matrix has an exact
 (doubly degenerate) eigenvalue
\begin{equation}
T(v)=(a(v)+b(v))^N=[\rho\Theta(0) h(v)]^N
\label{teigen}
\end{equation}
This is the eight vertex generalization of the six vertex
problem at
$\Delta=-1/2$ considered in \cite{strog},\cite{rs}, and \cite{dbnm}.

The TQ equation (\ref{tq}) for the eigenvalue of ${\bf T}(v)$ given by
(\ref{teigen}) is
\begin{equation}
h(v)^NQ_{N,\nu'}(v)=h(v-2K/3)^NQ_{N,\nu'}(v+4K/3)+h(v+2K/3)^NQ_{N,\nu'}(v-4K/3)
\label{help20}
\end{equation}
Using (\ref{p1}) and the fact that $N$ is odd (\ref{help20}) is rewritten as
\begin{equation}
h(v)^NQ_{N,\nu'}(v)
=-h(v+4K/3)^NQ_{N,\nu'}(v+4K/3)-h(v-4K/3)^NQ_{N,\nu'}(v-4K/3).
\label{tq2}
\end{equation}
It is easy to see that if $Q_{N,\nu'}(v)$ satisfies (\ref{tq2}) that
$Q_{N,\nu'}(-v)$ will satisfy the equation also. Thus we can require that
our two independent solutions satisfy
\begin{eqnarray}
Q_{N,\nu'}^e(-v)=Q_{N,\nu'}^e(v)\label{even}\\
Q_{N,\nu'}^o(-v)=-Q_{N,\nu'}^o(v)\label{odd}.
\end{eqnarray}

The odd solution, however, cannot occur. This can be seen if we note that both
$Q_{N,\nu'}^e(v)$ and $Q_{N,\nu'}^o(v)$ can be written as
\begin{equation}
Q_{N,\nu'}^{v_0}(v)=e^{-i \nu\pi v/2K}H(v-v_0)\prod_{j=1}^{(N-1)/2}H(v-v_j)H(v+v_j)
\end{equation}
with $v_j$ purely imaginary and an appropriate choice of $v_0$.
From the quasi-periodicity of $Q^{v_0}_{N,\nu'}$ it follows that the
sum rule holds (\ref{sumrule}) and thus we find that
\begin{equation}
{\rm Re}(v_0)=K
\end{equation}
and similarly the quasiperiodicity requires (\ref{nuform}) to hold and thus
 we find that either
\begin{eqnarray}
{\rm Im}(v_0)&=K',~~\nu'=0,~~\nu=1~~{\rm or}\label{nup0}\\
{\rm Im}(v_0)&=0,~~\nu'=1,~~\nu=0\label{nup1}
\end{eqnarray}
For either (\ref{nup0}) or (\ref{nup1}) it follows from the properties
(\ref{p1}) and (\ref{p2}) that for $v_0=0,iK'$ that
$Q^{v_0}_{N,\nu'}(v)$ is even
and thus $Q_{N,\nu'}^o(v)$ does not occur.

We now define
\begin{equation}
f_{N,\nu'}(v)=h(v)^NQ_{N,\nu'}(v)
\label{fdefn}
\end{equation}
and thus (\ref{tq2}) becomes
\begin{equation}
f_{N,\nu'}(v)+f_{N,\nu'}(v-4K/3)+f_{N,\nu'}(v+4K/3)=0.
\label{frecrel}
\end{equation}
From the quasiperiodicity properties
(\ref{nqp1}) and (\ref{nqp2}) and the evenness of $Q_{N,\nu'}(v)$ we see
that $f_{N,\nu'}(v)$ must also satisfy
\begin{eqnarray}
f_{N,\nu'}(v+2K)&=(-)^{1+\nu'}f_{N,\nu'}(v)\label{fp1}\\
f_{N,\nu'}(v+2iK')&=q^{-3N}e^{-3N\pi i v/K}f_{N,\nu'}(v)\label{fp2}\\
f_{N,\nu'}(-v)&=-f_{N,\nu'}(v)\label{fp3}
\end{eqnarray}

We satisfy (\ref{fp1}) and (\ref{fp3}) by writing
\begin{equation}
f_{N,\nu'}(v)=\sum_{j=-\infty}^{\infty}{\tilde f}_{j,\nu'}
e^{i\pi (j+{1+\nu'\over 2})v/K}
\end{equation}
with
\begin{equation}
{\tilde f}_{-j,\nu'}=-{\tilde f}_{j-1-\nu',\nu'}
\end{equation}
and find from (\ref{fp2}) that
\begin{equation}
{\tilde f}_{j+3N,\nu'}=q^{3N+2j+1+\nu'}{\tilde f}_{j,\nu'}.
\end{equation}
Thus defining
\begin{equation}
{a}_{j,\nu'}=q^{-(j+{1+\nu'\over 2})^2/3N}{\tilde f}_{j,\nu'}
\end{equation}
we have
\begin{equation}
{a}_{j+3N,\nu'}={a}_{j,\nu'},~~a_{-j',\nu'}=-a_{j-1-\nu', \nu'}
\label{period}
\end{equation}
and
\begin{equation}
f_{N,\nu'}(v)=\sum_{j=-\infty}^{\infty}a_{j,\nu'}q^{(j+{1+\nu'\over 2})^2/3N}
e^{i\pi (j+{1+\nu'\over 2})v/K}.
\label{help10}
\end{equation}
Setting $j=r3N+l$ and using (\ref{period}) in (\ref{help10})
we find
\begin{equation}
f_{N,\nu'}(v)=\sum_{l=0}^{3(N-1)/2}a_{l,\nu'}
\sum_{r=-\infty}^{\infty}q^{3N(r+{1+\nu'+2l\over 6N})^2}
[e^{i\pi (r+{1+\nu'+2l\over 6N})3Nv/K}-e^{-i\pi (r+{1+\nu'+2l\over 6N})3Nv/K}]
\label{help1}
\end{equation}
which is expressed in terms of the standard theta
function with characteristics
$\Theta{\epsilon \atopwithdelims[] \epsilon '}(v,\tau),$
whose definition and some useful properties are given in the appendix, as
\begin{equation}
f_{N,\nu'}(v)=\sum_{l=0}^{3(N-1)/2}a_{l,\nu'}
\Theta_o{(1+\nu'+2l)/3N \atopwithdelims[] 0}
(3Nv,3N\tau).
\label{fform}
\end{equation}


where
\begin{equation}
\Theta_o{(1+\nu'+2l)/3N\atopwithdelims[] 0}(3Nv,3N\tau)
= \Theta{(1+\nu'+2l)/3N\atopwithdelims[] 0}(3Nv,3N\tau)
-\Theta{(1+\nu'+2l)/3N\atopwithdelims[] 0}(-3Nv,3N\tau).
\end{equation}

We now use the form (\ref{fform}) in the difference equation (\ref{frecrel})
and then use (\ref{theta2}) to obtain
\begin{equation}
\sum_{l=0}^{3(N-1)/2}a_{l,\nu'}\{1+e^{i 2\pi(1+\nu'+2l)/3}+e^{-i 2\pi(1+\nu'+2l)/3}\}
\Theta_o{(1+\nu'+2l)/3N\atopwithdelims[] 0}(3Nv,3N\tau)=0
\end{equation}
This equation is satisfied if (for $\nu'=0,1$)
\begin{equation}
a_{l,\nu'}=0~~{\rm for}~~1+\nu'+2l\equiv 0~{\rm mod} 3.
\end{equation}
Then in (\ref{fform}) for $l\equiv 0~{\rm mod}3$ we use
(\ref{theta7}) to send the characteristic
$(1+\nu'-2l)/3N$ to $ -(1+\nu'-2l)/3N,$
and $a_{l,\nu'}\rightarrow -a_{l,\nu'}.$ Then if  we  use (\ref{theta6})
and (\ref{theta7}) and send $a_{l,\nu'}\rightarrow a_{l-(N-1)/2,\nu'}$
we may reorder the terms in (\ref{fform})  to write the
expression for $f_N(v)$ in the more elegant form
\begin{equation}
f_{N,\nu'}(v)=\sum_{l=-(N-1)/2}^{(N-1)/2}{\tilde a}_{l,\nu'}
\Theta_o{(6l-1-\nu')/3N \atopwithdelims[] 0}
(3Nv,3N\tau)
\label{fform2}
\end{equation}

To determine the remaining $N$ coefficients ${\tilde a}_{l,\nu'}$
we note that from the definition (\ref{fdefn}) that $f_{N,\nu'}(v)$ must have
an $Nth$ order zero at $v=0$ and $v=iK'$ and thus the first $N-1$
derivatives of $f_{N,\nu'}(v)$ must vanish at $v=0,~iK'.$

To satisfy the condition at $v=0$ we note that because $f_{N,\nu'}(v)$
is odd this requires that the odd derivatives
up to order $N-2$ must vanish and therefore we have the $(N-1)/2$ equations
\begin{equation}
\sum_{l=-(N-1)/2}^{(N-1)/2}{\tilde a}_{l,\nu'}
\Theta_o^{(2m-1)}{(6l-1-\nu')/3N \atopwithdelims[] 0}(0,3N\tau)=0
\label{lin1}
\end{equation}
for $m=1,2,\cdots, (N-1)/2$
where the superscript $2m-1$ indicates the $2m-1$ derivative with
respect to $v.$

To satisfy the condition at $v=iK'$ we note that for any $\epsilon$
\begin{equation}
\Theta_o{\epsilon\atopwithdelims[] 0}(3N(v+iK'),3N\tau)
=q^{-3N/4}e^{-\pi i3Nv/2K}
\Theta_o{1+\epsilon \atopwithdelims[] 0}(3Nv,3N\tau)
\end{equation}
Therefore $e^{\pi i3Nv/2K}f_{N,\nu'}(v+iK')$ is an odd function of $v$
which is given by
\begin{equation}
e^{\pi i3Nv/2K}f_{N,\nu'}(v+iK')
=q^{-3N/4}
\sum_{l=-(N-1)/2}^{(N-1)/2}{\tilde a}_{l,\nu'}
\Theta_o{1+(6l-1-\nu')/3N \atopwithdelims[] 0}(3Nv,3N\tau).
\label{oddf}
\end{equation}
This odd function has a zero of order $N$ at $v=0$ and thus
we find the companion equations to (\ref{lin1}) of
\begin{equation}
\sum_{l=-(N-1)/2}^{(N-1)/2}{\tilde a}_{l,\nu'}
\Theta_o^{(2m-1)}{1+(6l-1-\nu')/3N \atopwithdelims[] 0}(0,3N\tau)=0
\label{lin2}
\end{equation}
for $m=1,2,\cdots, (N-1)/2$

The $N$ coefficients ${\tilde a}_l$ of the expansion (\ref{fform2})
are thus determined from $N-1$ homogeneous equations (\ref{lin1}) and
(\ref{lin2}) and hence the existence of solutions for the TQ equation for
the eigenvalue (\ref{teigen}) of ${\bf T}(v)$ has been demonstrated.

Finally we show that
\begin{equation}
q^{3N/4}e^{\pi i3Nv/2K}f_{N,0}(v+iK')
=f_{N,1}(v)
\label{f1}
\end{equation}
from which if follows that
\begin{equation}
q^{3N/4}e^{\pi i3Nv/2K}Q_{N,\nu'=0}(v+iK')
=Q_{N,\nu'=1}(v).
\end{equation}

The result (\ref{f1}) follows from (\ref{oddf}) if we can show that
\begin{eqnarray}
\sum_{l=-(N-1)/2}^{(N-1)/2}{\tilde a}_{l,0}
\Theta_o{1+(6l-1)/3N \atopwithdelims[] 0}(3Nv,3N\tau)
=\sum_{l=-(N-1)/2}^{(N-1)/2}{\tilde a}_{l,1}
\Theta_o{(6l-2)/3N \atopwithdelims[] 0}(3Nv,3N\tau).
\end{eqnarray}
If $(N-1)/2$ is even this will follow if we can show that
\begin{eqnarray}
a_{l,0}=a_{(N-1)/2-l+1,1}~~{\rm for}~~1\leq l \leq (N-1)/2\label{f2}\\
a_{l,0}=a_{-l-(N-1)/2,1}~~{\rm for}~~-(N-1)/2 \leq l \leq 0
\label{f3}
\end{eqnarray}
and (\ref{f2}) and(\ref{f3}) are easily seen to follow
from (\ref{lin1}) and (\ref{lin2}).
The case $(N-1)/2$ odd is treated in a similar manner
and thus (\ref{f1}) is established.

The linear equations (\ref{lin1}) and (\ref{lin2}) can of course be
solved as determinants. It would be much more helpful, however, if these
determinants could be evaluated in some simpler form. Such a simplification,
if it exists, unfortunately involves identities in theta constants
which do not seem to be in the literature. We thus content ourselves
here with   the remark that for $N=3$ we have shown that the
coefficients ${\tilde a}_l$ are modular forms for the subgroup of the modular
group defined by the transformation
\begin{equation}
\tau\rightarrow {a\tau+b\over c\tau+d}
\end{equation}
where
\begin{equation}
a\equiv 1~{\rm mod}6N,~~b\equiv 0~{\rm mod}2,~~c\equiv~0~{\rm mod}3N,
~~d\equiv 1~{\rm mod}6N
\end{equation}

\section{The six vertex limit}

We may now consider limit $q\rightarrow 0$ where
the eight vertex model reduces to the six vertex model. In this limit
\begin{equation}
\eta\rightarrow {m\pi\over 2L}
\end{equation}
and in the XXZ Hamiltonian (\ref{hxxz}) we have
\begin{equation}
\Delta=\cos{m\pi\over L}
\end{equation}

\subsection{$N$ odd, $\eta=mK/L$ with $m$ odd and generic $\eta$}

In the eight vertex model  we saw in section 2
for $N$ odd, $\eta=mK/L$ with $m$ odd and in sec. 3 for generic $\eta$
that all states occur in pairs
such that for every set of roots $v_k$ there is a companion set of roots
$v_k+iK'$ and that both eigenvalues of $Q_{72}(v)$ correspond to the
same (doubly degenerate) eigenvalues of $T(v).$ This double
degeneracy of all transfer matrix eigenvalues for chains of odd length follows
from the spin inversion symmetry of the Boltzmann weights.

Consider the fundamental region to be given by
\begin{equation}
0\leq{\rm Re}(v)<K~~{\rm and}~~-K'/2\leq {\rm Im}(v)>3K'/2
\end{equation}
Then the pairing of solutions $v_k$ means that for every solution with
$n$ roots in the region
\begin{equation}
-K'/2\leq {\rm Im} (v)<K'/2
\label{lower}
\end{equation}
and $N-n$ roots in the region
\begin{equation}
K'/2\leq {\rm Im}(v)<3K'/2
\label{upper}
\end{equation}
there will be a corresponding solution with
$N-n$ roots in (\ref{lower}) and $n$ roots in (\ref{upper}).
Numerical studies indicate that as $q\rightarrow 0$ that the $n$ roots
in (\ref{lower}) stay a finite distance from $v=0$ and the $N-n$ roots in
(\ref{upper}) stay a finite distance from $v=iK'$. Therefore in the
limit $q\rightarrow 0$  with $v$ held fixed the quasi periodic functions
which are the eigenvalues of $Q_{72}(v)$ and which always
have $N$ zeroes in the fundamental region reduce to trigonometric
polynomials which have $n$ zeroes in the strip $0\leq {\rm Re}(v)<\pi$
where $n$ can take ALL values from $0$ to $N.$
These $n$ zeroes satisfy the Bethe's equation obtained from taking
the $q\rightarrow 0$ limit of (\ref{bethe3})
\begin{equation}
\left({\sin(v_l-m\pi/2L)\over\sin(v_l+m\pi/2L)}\right)^N=e^{2\pi i \nu m/L}
\prod_{{j=1\atop j\neq l}}^n{\sin(v_l-v_j-m\pi/L)\over \sin(v_l-v_j+m\pi/L)}.
\label{bethe5}
\end{equation}
We note in particular the following points:

1) To every solution of (\ref{bethe4}) with $n\leq (N-1)/2$ roots there
exists a ``companion'' solution with $N-n$ roots  which gives the same
eigenvalue of the transfer matrix $T(v).$ This is the
phenomenon discovered by Pronko and Stroganov \cite{ps}
in the case where $\gamma$ in the $XXZ$ Hamiltonian (\ref{hxxz})
is irrational. Examples of such pairs are given in table VI.

2) There are never any solutions with infinite roots or L strings
such as occur for even $N$ as seen in \cite{bax2002} and \cite{dfm}.

3)The limit of the Bethe equation (\ref{bethe2}) is formally the same as
the limit of (\ref{bethe3}) but the equation (\ref{bethe2}) does not allow
solutions with $n\geq (N+1)/2.$

This resolution of the question of the existence of solutions  to the
Bethe's equation (\ref{bethe5}) with $n>N/2$ with odd $N$
is very different from the
resolution for even $N$ presented in \cite{bax2002}
which involves both L strings and roots at infinity.

\begin{table}
\caption{ Roots of $Q(v)$ for the six vertex model with $\Delta=+1/2$ with
$N=9$ illustrating the pairing of solutions with $n$ and $N-n$ roots.
Each of the paired solutions corresponds to the same eigenvalue
of the six vertex transfer matrix and $XXZ$ spin chain.The energy
eigenvalue $E_{XXZ}$ of $H_{XXZ}$ and the corresponding
momentum $P$ are given  and the $\pm$ are to be taken independently.
The roots for the four negative values of $P$ are the complex conjugates of
roots for positive $P$}
\begin{tabular}{|l|l|l|l|}
\hline
$n=0$&$n=9$&$E_{XXZ}$&$P$\\ \hline\hline
none&$\pm i 0.74839\dots\cdots$&$-2.25$&$0$\\
&$0$&&\\
&$\pm i1.52761\cdots+\pi/2$&&\\
&$\pm i0.3.154723\cdots\pm1.047189\cdots$&&\\
\hline\hline
$n=1$&$n=8$& &\\ \hline \hline
0&$\pm i1.261399\cdots$&$-3.25\cdots$&$0$\\
&$\pm i 0.281540\cdots$&&\\
&$\pm i 0.636873\cdots\pm 1.048336\cdots $&&\\
\hline
$-i0.742104\cdots$&$i1.220370\cdots$ &$-2.78208\cdots$&$2\pi/9$\\
&$i0.228906\cdots$&&\\
&$-i0.361558\cdots$&&\\
&$-i1.401228\cdots$&&\\
&$ \pm i0.591172\cdots\pm 1.048234\cdots$&&\\ \hline
$-i0.844260\cdots+\pi/2$&$i1.163039\cdots$&$-1.59729\cdots$&$4\pi/9$\\
&$i0.183634\cdots$&&\\
&$-i0.616811\cdots$&&\\
&$-i0.737347\cdots +\pi/2$&&\\
&$\pm i 0.559024\cdots\pm 1.0480858\cdots$&&\\ \hline
$-i0.346573\cdots+\pi/2$&$i1.176586\cdots$&$-0.25$&$6\pi/9$\\
&$i0.126324\cdots$&&\\
&$-i1.060573$&&\\
&$-i0.359837\cdots+\pi/2$&&\\
&$\pm i0.528842\cdots\pm  1.047924\cdots$&&\\ \hline
$-i0.102156\cdots+\pi/2$&$i1.155571\cdots$&$0.629385\cdots$&$8\pi/9$\\
&$i0.0486462\cdots$&&\\
&$-i1.128327\cdots$&&\\
&$-i0.983179\cdots+\pi/2$&&\\
&$\pm i 0.4920739\cdots\pm  1.045727\cdots$&&\\ \hline
\end{tabular}
\end{table}
\subsection{$\eta=mK/L$ with $m$ even}

For the eight vertex model with  $N$ odd and $m$ even we saw
in sec. 4 that there are two types of solutions. Those with
$L$ strings with $n_L\geq 1$ and with $2n_B$ roots which occur in pairs
$v^B_k,~v^B_k+iK'$ and those with neither $L$ strings nor paired roots where
there are two solution for  $Q_{N,\nu'}(v)$ where the solutions for
$\nu'=0,1$ are paired under $v\rightarrow v+iK'.$

The first type of solution is of the form previously seen for $N$ even
and the six vertex limit of these solutions will in general
contain $L$ strings which would seem to be in contradiction with
the statement of sec. 4 of \cite{bax2002} that ``for $N$ odd
no problems appeared'' in taking the limit $H\rightarrow 0.$

The second sort of solution will behave in the six vertex limit in a
fashion similar to the case of $m$ odd which is illustrated by
considering the limit $q\rightarrow 0$ in the solution for $f_{N,\nu'}(v).$
When $q\rightarrow 0$ we use (\ref{theta12}) to
find from (\ref{fform2})
that
\begin{equation}
{\rm lim}_{q\rightarrow 0}f_{N,\nu'}(v) = \sum_{l=-(N-1)/2}^{(N-1)/2}
b_{l,\nu'}2i\sin(6l-1-\nu')v
\label{fform26}
\end{equation}
where
\begin{equation}
b_{l,\nu'}={\rm lim}_{q\rightarrow 0}{\tilde a}_{l,\nu'}q^{(6l-1-\nu')^2/36N^2}
\end{equation}
and from (\ref{lin1}) the  $b_{l,\nu'}$ satisfy
\begin{equation}
\sum_{l=-(N-1)/2}^{(N-1)/2}b_{l,\nu'}(6l-1-\nu')^{2m-1}
=0.
\label{lin16}
\end{equation}
For the remaining equation (\ref{lin2})  we use again (\ref{theta12})
to find for small $q$ that
\begin{eqnarray}
&\sum_{l=-(N-1)/2}^0b_{l,\nu'}q^{[3N+2(6l-1-\nu')]/12N}
(3N+6l-1-\nu')^{2m-1}\nonumber\\
&+\sum_{l=1}^{(N-1)/2}b_{l,\nu'}q^{[3N-2(6l-1-\nu')]/12N}
(-3N+6l-1-\nu')^{2m-1}=0.
\label{lin26}
\end{eqnarray}
We note that in (\ref{lin26}) there are two types of terms: those where the
coefficient of $b_{l,\nu'}$ vanishes as $q\rightarrow 0$ which impose
no constraint on $b_{l,\nu'}$ and those where the coefficient of
$b_{l,\nu'}$ diverges as $q\rightarrow 0$ which forces the $b_{l,\nu'}$
to vanish. We  thus find

{\bf for $\nu'=0$}

\begin{equation}
{\rm for}~{N-1\over 4}~{\rm integer}~~
b_{l,0}\neq 0~~{\rm only~for}~-{N-1\over 4}\leq l \leq {N-1\over 4}
\label{case1}
\end{equation}

and

\begin{equation}
{\rm for}~{N-3\over 4}~{\rm integer}~~
b_{l,0}\neq 0~~{\rm only~for}~-{N-3\over 4}\leq l \leq {N-3\over 4}
\label{case2}
\end{equation}

while

{\bf for $\nu'=1$}

\begin{equation}
{\rm for}~{N-1\over 4}~{\rm integer}~~
b_{l,0}\neq 0~~{\rm only~for}~-{N-1\over 4}+1\leq l \leq {N-1\over 4}
\label{case3}
\end{equation}

and

\begin{equation}
{\rm for}~{N-3\over 4}~{\rm integer}~~
b_{l,0}\neq 0~~{\rm only~for}~-{N-3\over 4}\leq l \leq {N-3\over 4}+1
\label{case4}
\end{equation}

For $\nu'=0$ and $(N-1)/4$ an integer
we use (\ref{case1}) in (\ref{fform26}) and (\ref{lin16}) with
$l=(N-1)/4-k$ and $b_{l,0}=\alpha_k$ to find
\begin{equation}
{\rm lim}_{q\rightarrow 0}f_{N,0}(v)=-2i\sum_{k=0}^{(N-1)/2}
\alpha_k \sin (6k-3(N-1)/2+1)
\label{strog1}
\end{equation}
with
\begin{equation}
\sum_{k=0}^{(N-1)/2}\alpha_k(6k-3(N-1)/2+1)^{2m-1}=0
\label{strog2}
\end{equation}
and for $\nu'=1$ and $(N-3)/4$ an integer we use (\ref{case4}) in
(\ref{fform26}) and (\ref{lin16}) with
$l=k-(N-3)/4$ and $b_{l,0}=\alpha_k$ to find
\begin{equation}
{\rm lim}_{q\rightarrow 0}f_{N,1}(v)=2i\sum_{k=0}^{(N-1)/2}
\alpha_k \sin (6k-3(N-1)/2+1)
\label{strog3}
\end{equation}
where again $\alpha_k$ satisfies (\ref{strog2}).

For the 6 vertex model the problem of finding a solution
of the TQ
equation with $(N-1)/2$ Bethe roots for the 6 vertex eigenvalue
$T_6(v)=(\sin v)^N$ has been previously studied by Stroganov \cite{strog}.
If we note that the function $f_{\rm strog}(v)$ of \cite{strog} satisfies
\begin{equation}
 f_{\rm strog}(v+\pi)=(-1)^{(N+1)/2}f_{\rm strog}(v)
\end{equation}
instead of (\ref{fp1}) it is seen that (\ref{strog1})-(\ref{strog3}) agrees with
$f_{\rm strog}(v)$ for all odd $N.$

For $\nu'=0$ and $(N-3)/4$ an integer  we use (\ref{case2}) in
(\ref{fform26}) and (\ref{lin16}) with $l=k-(N-3)/4$
and $b_{l,0}=\beta_k$ to find
\begin{equation}
{\rm lim}_{q\rightarrow 0}f_{N,0}(v)=2i\sum_{k=0}^{(N-3)/2}
\beta_k \sin (6k-3(N-1)/2+2)
\label{strog4}
\end{equation}
where
\begin{equation}
\sum_{k=0}^{(N-3)/2}
\beta_k (6k-3(N-1)/2+2)^{2m-1}=0
\label{strog5}
\end{equation}
and for $\nu'=1$ and $(N-1)/4$ an integer we use (\ref{case3}) with
$l=(N-1)/4-k$ and $\beta_k=b_{l,1}$ to find
\begin{equation}
{\rm lim}_{q\rightarrow 0}f_{N,1}(v)=2i\sum_{k=0}^{(N-3)/2}
\beta_k \sin (6k-3(N-1)/2+2)
\label{strog6}
\end{equation}
where the $\beta_k$ still satisfy (\ref{strog5}). These functions are
of the form of the ``second solution'' g(v)  eqn. (12) of
Stroganov \cite{strog} except the upper limit of (\ref{strog4}) and
(\ref{strog6}) is $(N-3)/2$ instead of $(N-1)/2$

\section{Discussion}

The difficulty of discussing what in the literature  is called ``Baxter's Q''
is that over the years three different
objects have been defined which have all been denoted by 
the same symbol and all of which
satisfy a ``TQ''  equation (\ref{tq}).

\subsection{Three definitions of $Q$}

The first definition of  Q is the matrix,
which we have here called ${\bf Q}_{72}(v),$ defined in \cite{bax72}
only when $\eta$ satisfies the root of unity condition (\ref{root72}).
The definition requires an auxiliary matrix ${\bf Q}_{72,R}(v)$
to be nonsingular
and in paper I \cite{fm} we found that if  $m$ is even
and $N$ is odd or when $m$ and $N$ are both even and $N\geq L-1$
then the nonsingularity assumption on ${\bf Q}_{72,R}(v)$ fails.
The matrix ${\bf Q}_{72}(v)$ commutes with ${\bf S}$ but not with $\bf{R}$
and satisfies the quasiperiodicity relations (\ref{qp1}) and (\ref{qp2}).
We have seen repeatedly in \cite{fm} and in this present paper that
the eigenvalues of ${\bf Q}_{72}(v)$ are in general of the form
(\ref{fac3})
\begin{equation}
Q_{72}(v)={\cal K}(q,v_k){\rm exp}(-i\nu \pi v/2K)\prod_{j=1}^NH(v-v_j)
\label{ev72}
\end{equation}
where the $v_j$ satisfy the sum rules (\ref{sumrule}) and (\ref{nuform}).
The matrix ${\bf Q}_{72}(v)$ has no degenerate
eigenvalues even though many of the eigenvalues of ${\bf T}(v)$ are degenerate.

The second definition of  Q is the matrix we have here
called ${\bf Q}_{73}(v)$
defined in sec. 6 of \cite{bax731}
and in sec. 10.5 of \cite{baxb}
for generic values of $\eta$ not just those satisfying
the root of unity condition (\ref{root72}). This definition applies
only for $N$ even. The matrix ${\bf Q}_{73}(v)$ commutes with
both $\bf S$  and $\bf R$ and it is shown in (10.5.43) of \cite{baxb} that
it satisfies the quasiperiodicity relations (\ref{qp731} and (\ref{qp732}).
All eigenvalues of ${\bf Q}_{73}(v)$ are of the form (123) of \cite{bax2002}
[and (10.6.8) of \cite{baxb}]
\begin{equation}
Q_{73}(v)=e^{2i\tau v}\prod_{j=1}^{N/2}h(v-v_j)
\label{ev73}
\end{equation}
where the $v_j$ satisfy the sum rule (125) of \cite{bax2002}
[ and (10.6.7) of \cite{baxb}] which depends on
the eigenvalues of $\bf S$ and $\bf R.$
The matrices ${\bf Q}_{72}(v)$ and ${\bf Q}_{73}(v)$ are
not similar to each other and the  eigenvalues of the
form (\ref{ev73}) with generic $\eta$
cannot in general specialize to (\ref{ev72}) when $\eta$ is a root of unity.

There is yet one more definition of Q given in (1.24) of \cite{bax733}
and (132) of \cite{bax2002}. In this definition
$Q(v)$ is (a set of) scalar function(s) defined by
\begin{equation}
Q(v)=\prod_{j=1}^{n_B} h(v-v_j)
\label{evb}
\end{equation}
where [see (\ref{cond73})] $N=2n_B+L_{73}n_L,$
the $v_j$ are the solutions of the Bethe's equation (\ref{bethe2})
which  is derived for the related solid on solid
model defined for $\eta$ satisfying the root of unity
condition (\ref{root72})
and there is no sum rule on the $v_j$ unless
$n_L=0$ [see footnote 15 of \cite{bax2002}].
The properties of this scalar $Q(v)$ under translations
$v\rightarrow v+2K$ and $v \rightarrow v+iK'$ [given in (134) of
\cite{bax2002}] are obtained directly from (\ref{thtrans}) and in contrast
with the quasiperiodicity properties (\ref{qp731}) and (\ref{qp732}) will
explicitly involve the sum of the roots $\sum_k v_k$ which cannot
be eliminated due to the lack of a sum rule.

From the definition (\ref{evb}) of $Q(v)$ a matrix
is constructed in (1.29) of \cite{bax733} with the assumption that
all eigenvalues of ${\bf T}(v)$ can be obtained from these scalar functions
$Q(v)$ which are now regarded as eigenvalues of some matrix.
However all eigenvalues of ${\bf T}(v)$
for odd $N$ and many eigenvalues for even $N$ are at least doubly degenerate.
All degenerate eigenvalues of
${\bf T}(v)$ have the same $n_B$ Bethe roots $v_j$ in (\ref{evb})
and thus this construction
must lead to a matrix ${\bf Q}(v)$ which also has degenerate
eigenvalues and hence the eigenvalues of this ${\bf Q}(v)$ contain
no information about the multiplicities of the eigenvalues
of ${\bf T}(v).$

\subsection{Quasiperiodicity conditions}

All three definitions  of $Q(v)$ discussed above solve the same TQ equation.
They differ only in which quasiperiodicity conditions
(\ref{nqp1})-(\ref{nqp2}) or (\ref{qp731})-(\ref{qp732}) are imposed.
These may be thought of as boundary conditions for the TQ difference equation.
We emphasize that at roots of unity the TQ equation is in general
not sufficient to determine all the eigenvalues of ${\bf Q}(v)$ but is it
a necessary condition that any ${\bf Q}(v)$ matrix must satisfy.

The quasiperiodicity condition (\ref{nqp1})-(\ref{nqp2})
comes from the assumption
that ${\bf Q}(v)$ commutes with ${\bf S}$ and that all eigenvalues of
${\bf Q}(v)$ are quasiperiodic entire functions with $N$ zeroes in the
fundamental region of the Boltzmann weights (\ref{bw}).
If ${\bf Q}(v)$ has no degenerate eigenvalues and if its rank
is $2^N$ then the eigenvalues of ${\bf Q}(v)$ will characterize all
the (possibly degenerate) eigenvalues of ${\bf T}(v).$

The quasiperiodicity
condition (\ref{qp731})-(\ref{qp732}) imposes the additional restriction
that the zeroes must occur in pairs $v_k$ and $v_k+iK'$ (which is
possible only for $N$ even).
These quasiperiodicity conditions  have been
extensively discussed in the literature. They are the conditions used in
\cite{bax731}-\cite{bax733}, \cite{bax2002} and \cite{baxb} to solve the
8 vertex model with $N$ even for $\eta$ either generic or a root of unity.
A mathematical study of the solutions of the TQ equation ({\ref{tq})
with these quasiperiodicity conditions was made in \cite{kric}.

The principle ``new development'' of this and the previous paper \cite{fm}
is the introduction of the quasiperiodicity
conditions (\ref{nqp1})-(\ref{nqp2}) which are less restrictive than
the quasiperiodicity conditions (\ref{qp731}) and (\ref{qp732})
satisfied by ${\bf Q}_{73}(v).$

The quasiperiodicity condition (\ref{nqp2}) was first seen  in \cite{fm}
to apply to ${\bf Q}_{72}(v)$ defined for roots of unity.
For odd $N$ (\ref{nqp2}) must be used for all
$\eta$ because for  odd $N$ ${\bf R}$
and ${\bf S}$ do not commute and it is impossible to diagonalize
${\bf Q}(v),~{\bf S}$ and ${\bf R}$ at the same time. Hence it follows that
(\ref{qp732}) cannot be applied in this case.

\subsection{Second solutions}

The TQ equation is a second order difference equation and as such it
is expected to have two independent solutions.
For the quasiperiodicity conditions (\ref{nqp1})-(\ref{nqp2})
this is quite
obvious because if  ${\bf Q}(v)$ satisfies the TQ equation (\ref{tq})
then $e^{i \pi Nv/2K}{\bf Q}(v+iK')$ satisfies the same equation.
Therefore as long as  ${\bf Q}(v)$ and $e^{i \pi Nv/2K}{\bf Q}(v+iK')$
are linearly independent the TQ
(\ref{tq}) equation will have two solutions. For odd $N$ these two
solutions  are characterized by the two values of quantum number $\nu'=0,1$
whereas for even $N$ both solutions have the same value of $\nu'.$
These two linearly independent solutions for ${\bf Q}(v)$ correspond to
the  degeneracy of the eigenvalues of ${\bf T}(v)$ under spin reversal.

The only situation where this linear independence fails is when
the quasiperiodicity (\ref{qp732}) holds. This can only happen for
$N$ even and  the mathematical existence of a
second solution for $\eta$ not a root of unity has been discussed
in \cite{kric}.
This solution is the analogue of a second solution of linear differential
equations with equal roots of the indicial equation. These solutions will
not be entire functions of the variable $v$ and hence will not be
allowed solutions for the 8 vertex model. This agrees with
fact that for $N$ even and $\eta$ not a root of unity the eigenvalues of
${\bf T}(v)$ are not degenerate \cite{bax731}-\cite{bax733}.

\subsection{The six vertex limit}

For odd $N$ the six vertex limit of the TQ equation with the
quasiperiodicity conditions (\ref{nqp2}) is easily taken. For
$m$ odd the two
independent solutions ${\bf Q}(v)$ and $e^{i\pi N v/2K}{\bf Q}(v+iK')$
smoothly go to the solutions given by \cite{bax2002}, \cite{ps} and
\cite{korff} where one solution has   less that $N/2$ roots and one
solution has more that $N/2$ roots. The corresponding eigenvalues
of the transfer matrix are doubly degenerate under spin inversion.
The six vertex case for $m$ even is treated in \cite{korff}.

For even $N$ the limit is more subtle.
When $\eta$ is not a root of unity and $N$ is even
the quasiperiodicity relation (\ref{qp732}) holds. This quasiperiodicity
relation does not
allow for spin doublets.
The eigenvectors of ${\bf Q}(v)$ are eigenvectors of $\bf R$ and the
eigenvalues which correspond to the two different eigenvalues of $\bf R$
are different \cite{bax731}-\cite{bax733}, \cite{baxb}. In the
6 vertex limit the eigenvectors of the 8 vertex model will go to linear
combinations of Bethe vectors which are eigenvectors of $S^z.$
For generic $\eta$ the six vertex model has two solutions \cite{ps}
${\bf Q}(v)$  of the $TQ$ equation but one of these violates the
analyticity assumptions needed for ${\bf Q}(v)$ to be relevant for
the 6 vertex model. A resolution of this in terms of roots at infinity
is given in \cite{bax2002}.

The case of even  $N$ and $\eta$ a root of unity is also discussed in
\cite{bax2002} and \cite{korff}.

\subsection{Conclusion}

In this paper we have seen for odd $N$ with $\eta=mK/L$ and $m$ odd
that the zeroes of ${\bf Q}_{72}(v)$ do not satisfy what is usually
called ``Bethe's'' equation (\ref{bethe2}) but instead satisfy the more general
generic equation for roots (\ref{bethe4}). We conjecture that  
for irrational $\eta/K$ a matrix ${\bf Q}(v)$ exists which satisfies
the $TQ$ equation (\ref{tq}) with (\ref{tqcom}), ({\ref{qqcom})and
the quasiperiodicity conditions (\ref{nqp1}), (\ref{nqp2}). The roots
of this conjectured matrix will
also satisfy ({\ref{bethe4}). For neither $\eta=mK/L$ with $m$ odd nor 
for irrational $\eta/K$ will there be any $L$ strings or paired roots
$v_k,~v_k+iK.'$

For $N$ odd and $\eta=2mK/L$ no ${\bf Q}(v)$ is known to exist but the
limit $\eta\rightarrow 2mK/L$ was studied above and it was seen that
there are two distinct classes of eigenvalues $Q(v)$: 1) those which
do not have $L$ strings or paired roots and 2) those which have  an odd
number of $L$ strings $n_L\geq 1$ and the remaining roots all occur
in pairs. This second class of eigenvalues are obtained from the
eigenvectors of $T(v)$ computed in ref. \cite{bax731}-\cite{bax733}
from considering the related SOS model. We conjecture that
a ${\bf Q}(v)$ matrix exists which will incorporate both the states
computed by ref.\cite{bax731}-\cite{bax733} and those which are not
computed by those methods. The existence of $L$ strings and degenerate
eigenvalues of ${\bf T}(v)$ implies that this ${\bf Q}(v)$ will not be
unique. 

\vspace{.2in}

{\Large \bf Acknowledgments}

We wish to thank Profs. H. Farkas, I. Kra, and  T. Miwa  
for fruitful discussions.
One of us (BMM) is pleased to thank Prof. M. Kashiwara
for hospitality at the Research Institute of Mathematical
Sciences of Kyoto University where part of this work was carried out.
This work is partially supported by NSF grant DMR0302758.

\vspace{.2in}

\appendix

\section{Properties of theta functions}

The definition of Jacobi theta functions of nome $q$ is
\begin{equation}
H(v)=2\sum_{n=1}^{\infty}(-1)^{n-1}q^{(n-{1\over 2})^2}
\sin [(2n-1)\pi v/2K]
\end{equation}
\begin{equation}
\Theta(v)=1+2\sum_{n=1}^{\infty}(-1)^nq^{n^2}\cos (nv\pi/K)
\label{thetadef}
\end{equation}
where $K$ and $K'$ are the standard elliptic integrals of the first kind and
\begin{equation}
q=e^{-\pi K'/K}=e^{i\pi\tau}
\end{equation}
These theta functions satisfy the quasiperiodicity relations
\begin{eqnarray}
H(v+2K)&=-H(v)\label{p1}\\
H(v+2iK')&=-q^{-1}e^{-\pi i v/K}H(v)\label{p2}\\
\Theta(v+2K)&=\Theta(v)\label{p3}\\
\Theta(v+2iK')&=-q^{-1}e^{-\pi i v/K}\Theta(v)\label{p4}
\end{eqnarray}
From (\ref{thetadef}) we see that $\Theta(v)$ and $H(v)$
are not independent but satisfy
\begin{eqnarray}
\Theta(v\pm iK')&=\pm iq^{-1/4}e^{\mp{\pi i v\over 2 K}}H(v)\nonumber\\
H(v\pm iK')&=\pm iq^{-1/4}e^{\mp{\pi i v\over 2 K}}\Theta(v)
\label{thtrans}
\end{eqnarray}

Theta functions with characteristics $\epsilon$ and $\epsilon '$ are defined
as
\begin{eqnarray}
\Theta{ \epsilon \atopwithdelims[] \epsilon '}(v,\tau)
&=\sum_{n=-\infty}^{\infty}e^{i \pi(n+\epsilon/2)^2\tau}
e^{\pi i (n+\epsilon/2)(u/K+\epsilon')}\nonumber\\
&=\sum_{n=-\infty}^{\infty}q^{(n+\epsilon/2)^2}
e^{\pi i (n+\epsilon/2)(u/K+\epsilon')}
\label{theta1}
\end{eqnarray}
In this notation
\begin{equation}
H(v)=-\Theta{1 \atopwithdelims[] 1}(v),~~ \Theta(v)=
\Theta{0 \atopwithdelims[] 1}
\end{equation}

These theta functions satisfy the quasiperiodicity conditions
\begin{equation}
\Theta{\epsilon \atopwithdelims[] \epsilon '}(v+2K,\tau)
=e^{i \pi \epsilon}
\Theta{\epsilon \atopwithdelims[] \epsilon '}(v,\tau)
\label{theta2}
\end{equation}
and
\begin{equation}
\Theta{\epsilon \atopwithdelims[] \epsilon '}(v+2iK',\tau)
=e^{\pi i \epsilon '}q^{-1}e^{-i \pi u/K}
\Theta{\epsilon \atopwithdelims[] \epsilon '}(v,\tau)
\label{theta3}
\end{equation}
These quasiperiodicity properties guarantee that
$\Theta{ \epsilon \atopwithdelims[] \epsilon '}(v,\tau)$ has exactly
one zero in the fundamental region
\begin{equation}
0\leq {\rm Re}(v) <2K,~~0\leq {\rm Im}(v)<2K'.
\label{fr}
\end{equation}
The functions have the further properties that
\begin{equation}
\Theta{\epsilon\pm2m\atopwithdelims[] \epsilon '\pm 2n}(v,\tau)
=e^{-i\pi n\epsilon}\Theta{\epsilon \atopwithdelims[] \epsilon '}(v,\tau)
\label{theta6}
\end{equation}

\begin{equation}
\Theta{-\epsilon \atopwithdelims[] -\epsilon'}(v,\tau)
=\Theta{\epsilon \atopwithdelims[] \epsilon '}(-v,\tau).
\label{theta7},
\end{equation}
\begin{equation}
\Theta{\epsilon \atopwithdelims[] \epsilon '}(v+iK')
=q^{-1/4}e^{-i\pi \epsilon'/2}e^{-i\pi v/2K}
\Theta{\epsilon +1\atopwithdelims[] \epsilon '}(v,\tau)
\label{theta9}
\end{equation}

and as $q\rightarrow 0$ with $|\epsilon|<1$ we have
\begin{equation}
\Theta{\epsilon \atopwithdelims[] \epsilon '}(v,\tau)\rightarrow
q^{\epsilon ^2/4}e^{\pi i {\epsilon \over 2}(v/K+\epsilon ')}.
\label{theta12}
\end{equation}

Further discussion and properties can be
found in reference books on theta functions such as \cite{k} and \cite{fk}.

\end{document}